\documentclass[twocolumn,showpacs,prl,amsmath,amssymb]{revtex4}
\usepackage{graphicx}
\usepackage{bm}
\begin{document}

\title{Quantum Information Protectorates in Coupled Quantum Dot
  Exchange Gates}
\author{V.W. Scarola and S. Das Sarma}
\affiliation{Condensed Matter Theory Center, 
Department of Physics, University of Maryland,
College Park, MD 20742-4111}

\begin{abstract}

Using exact diagonalization we study the low energy 
Hilbert space of the two-electron, two-quantum dot 
artificial molecule under a perpendicular magnetic field.  
We show that electrons bind to vortices to induce several 
spin transitions among ground states.  
Furthermore, the lowest excited states of either 
even or odd vorticity mix, opening an
anticrossing which protects the 
quantum information stored in the spin states 
of the strongly correlated quantum dot molecule.  

\end{abstract}
\pacs{}
\maketitle

Exchange gate, the key idea \cite{Loss} underlying spin 
quantum computation in semiconductor quantum dots, involves 
the tunable exchange coupling between two electrons localized 
in neighboring quantum dots.  This system can be thought of 
as a two-electron quantum molecule (i.e. an artificial 
${\text H}_2$ molecule.) in an external magnetic field with external gates 
controlling the ``molecular'' coupling between the two 
quantum dots.  In this Letter we show that the 
effective exchange coupling in such an artificial quantum 
dot molecule manifests highly non-trivial and unexpected 
magnetic field dependence which can be used to ``protect'' 
quantum information.

The Heisenberg model captures the essential, low energy  
spin physics of neighboring, single electron quantum 
dots \cite{Loss}.  For well separated and therefore weakly coupled 
electrons in the perturbative regime, recent studies \cite{Loss2,Hu}
indicate that such an artificial two-dot molecule 
may indeed be thought of 
as a two-level spin system with an effective exchange 
interaction.  However, there is no direct experimental 
probe of the average inter-electron distance or confinement.  
Bearing this in mind 
we study the nature of the ground and excited states of the 
strongly coupled, two electron system using both exact 
diagonalization and variational techniques 
in a regime inaccessible to perturbation theory.  

We first solve the problem of two electrons in one 
parabolic dot in a large, perpendicular 
magnetic field analytically.  We find 
ground and excited state transitions as a 
function of magnetic field.     
At special magnetic fields correlations force 
the excited states to 
become degenerate potentially 
destroying the two-level approximation 
invoked in applying the Heisenberg model
to two laterally separated dots.  
To explore this possibility in detail we examine
the two electron-two dot problem 
using exact diagonalization.
We find, as in a previous study \cite{Harju},  
striking evidence for $several$ 
spin transitions as a function of 
magnetic field.  The two lowest states still map onto 
a Heisenberg model but with an exchange interaction 
which oscillates with magnetic field, 
clearly a non-trivial, Coulombic effect.  
As for the higher excited states, we find that 
rotational symmetry breaking 
creates an anticrossing which, 
with the correct parameters, can 
be used to protect the quantum information 
stored in the two electron entangled state. 
We explore the nature of the states making 
up the anticrossing.
We show conclusively that trial states based on the 
composite fermion theory of the fractional 
quantum Hall effect 
accurately capture the 
ground and excited states.     
We classify these states according to the number of vortices attached to 
each electron, or vorticity.
Our principal conclusion is twofold:  Ground state 
spin transitions occur as a result of a unit increase in 
vorticity, and the lowest excited states of either 
even or odd vorticity mix, opening an energy gap which acts 
as a protectorate for quantum information.  

We begin with a general Hamiltonian describing two,  
lateral, single electron quantum dots:
\begin{eqnarray}
H(\omega_0, R)&=& \sum_{i=1}^{2}\left[ \frac{1}{2m^*}
\left( \textbf p_i+e\textbf A_i\right)^2 +V(\omega_0,R;\textbf r_i) \right]
\nonumber
\\ 
&+& \frac{e^2}{4\pi\varepsilon \vert \textbf r_1-\textbf r_2 \vert}
+g^*\mu_B \textbf S\cdot\textbf B. 
\label{FullH}
\end{eqnarray}
For  
GaAs we take the effective mass to be $m^*=0.067 m_\textrm{e}$ and 
the dielectric constant $\varepsilon=12.4\varepsilon_0$.  
The two dimensional coordinates $\textbf r=(x,y)$ lie in the plane 
perpendicular to the magnetic field, $B$, which points 
along the $z$-axis.  In the symmetric gauge we have: 
$\textbf A=\frac{B}{2}(-y,x,0)$.  The confinement 
potential consists of two parabolas separated by a 
distance $R$ along the $x$-axis:
\begin{eqnarray}
V(\omega_0,R;\textbf r)&=&\frac{m^*\omega_0^2}{2}\min\left\{
\left(x-\frac{R}{2}\right)^2+y^2,\right.
\nonumber
\\
&&\qquad\qquad\quad\ \left.\left(x+\frac{R}{2}\right)^2+y^2\right\}, 
\end{eqnarray}
where we choose the confinement parameter to be $\hbar\omega_0=3$ meV.  
The 
last term in $H$ is the Zeeman contribution where 
$\textbf S$ is the total electron spin and 
$g^*$ is the $g$-factor in GaAs.  In GaAs the Zeeman 
splitting, $\sim -0.025S_zB[$T$]$ meV, 
is an additive constant 
which may be taken into account after solving for 
the orbital degrees of freedom.  We may therefore restrict our 
attention to the $S_z=0$ subspace with no loss of generality.  
We seek solutions of the above 
Hamiltonian of the form $A[\psi \chi]$ where $\psi$ and $\chi$ 
are the orbital and spin parts of the wave function and 
$A$ is the antisymmetrization operator.  

The above model is analytically soluble in two extreme regimes:  
Two well-separated one-electron ``artificial atoms'' and a 
two-electron artificial atom in a high magnetic field. 
The first case is trivial and consists of two well separated 
quantum dots (akin to two well-separated one electron atoms $not$ 
in a molecular state) with one electron in each dot, $R\gg a$.  Here 
$a=\sqrt{\hbar/eB}(1+4\omega_0^2/\omega_c^2)^{-\frac{1}{4}}$ 
is a modified magnetic length and $\omega_c=eB/m^*$ is the 
cyclotron frequency.  In this case we may ignore the Coulomb 
interaction.  The non-interacting ground state consists of 
degenerate singlet and triplet states.  

In the second soluble limit (a two-electron artificial atom) 
two electrons lie in one 
parabolic dot in a strong magnetic field.  In this case 
we take $\omega_0 \ll \omega_c$ and $R=0$.  Correspondingly, 
the relative and $z$-component of angular momentum 
commute with the Hamiltonian.
At large magnetic fields we may project onto the lowest Landau 
level (LLL), giving:  
\begin{eqnarray}
H(\omega_0,R)\left\vert_{\omega_0\ll\omega_c,R=0}\right.
=\gamma\hat{L}_z
+\sum_{m=0}^{\infty}V_m \hat{P}_{m},
\label{LLLH}
\end{eqnarray}
where $\gamma\equiv\frac{\hbar}{2}
(\sqrt{\omega_c^2+4\omega_0^2}-\omega_c)$ and  
$\hat{L}_z$ is the total angular momentum    
in the $z$ direction.  The second term represents 
the LLL Coulomb interaction, projected 
onto eigenstates of relative angular momentum, $m$, 
via the projection operator $\hat{P}_{m}$.  The 
coefficients, $V_m$, are the Haldane pseudopotentials 
\cite{Haldane} which, for the Coulomb interaction, 
decrease with increasing $m$, at large $m$.  The unnormalized 
eigenstates of relative angular momentum are:
\begin{eqnarray}
\vert m\rangle=(z_1-z_2)^{m}
\exp\left(\frac{-\vert z_1\vert^2-\vert z_2\vert^2}{4a^2}\right),
\end{eqnarray}
where $z=x+iy$.  It can be shown by direct calculation 
that, because there is no center of mass motion, 
the above wave functions are also eigenstates 
of $\hat{L}_z$, with eigenvalue $m$.  Thus the 
set of states $\vert m\rangle$ form 
an orthogonal set of eigenstates of 
Eq.~(\ref{LLLH}), with eigenvalues 
$E_m=\gamma m+ V_m$.  The relative angular momentum 
of the lowest energy state depends on the parameters in 
$\gamma$ and the form of the interaction.  For the 
LLL Coulomb interaction,
in the artificial zero field limit, the lowest energy 
state has $m=1$.  Increasing $B$ lowers the confinement 
energy cost, $\gamma m\sim m/B$, and raises 
the Coulomb cost, $V_m \sim \sqrt{B}/(m+1)$, thereby raising $m$ by one.  
The transition from one eigenstate to the next occurs 
when $E_m=E_{m+1}$ which, for $\omega_0\ll\omega_c$, occurs 
at magnetic fields:
\begin{eqnarray} 
B_m\approx\left(\frac{C}{\tilde{V}_{m}-\tilde{V}_{m+1}} \right)^{\frac{2}{3}},
\end{eqnarray}
where $\tilde{V}_{m}\equiv V_m/(e^2/4\pi\varepsilon a)$ and 
$C\equiv4\pi\varepsilon\hbar^{3/2}\omega_0^2 m^*  e^{-7/2}$.
For the parameters used here we find $C\sim1.2 \text{ T}^{3/2}$.
The states $\vert m \rangle$ are symmetric (antisymmetric) 
with respect to 
particle exchange if $m$ is even (odd).  
The total wave function, $A[\psi\chi]$, must 
be antisymmetric.  Therefore $\chi$ is spin singlet (triplet) 
for $m$ even (odd).  Here, the index $m$ may be interpreted 
as the number of zeros or vortices attached to each electron, 
allowing us to assign a vorticity to each spin state. 
We show that vorticity plays an important role in the protection 
of quantum information in the GaAs coupled quantum dot exchange 
gate architecture.  

\begin{figure}
\includegraphics[width=2.6in]{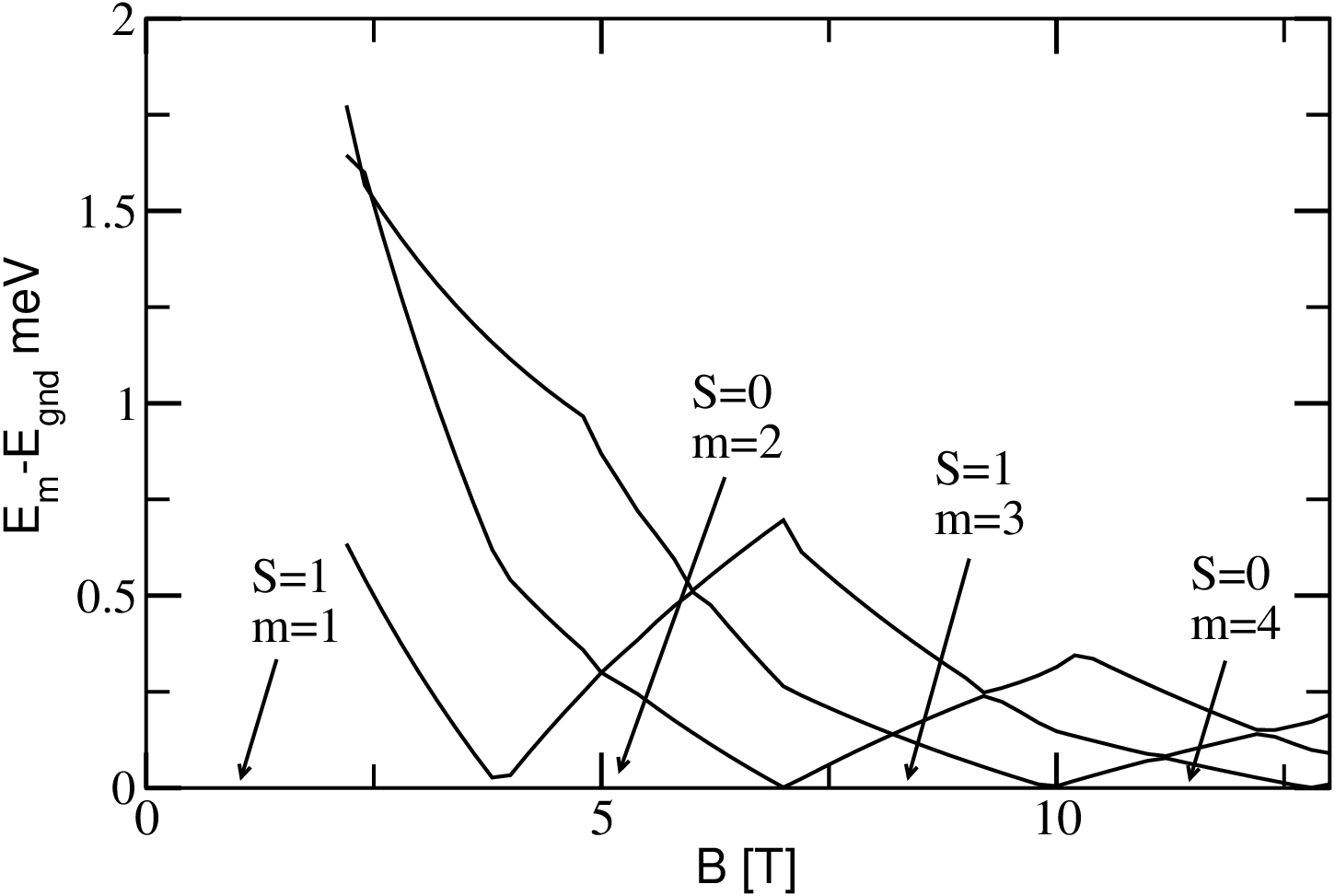}
\caption{
Energy of the 
four lowest states of a two electron quantum dot under a 
perpendicular magnetic field in the lowest Landau level plotted 
as a function of perpendicular magnetic field.  
The ground state energy is set 
to zero.  The ground state alternates between spin singlet $(S=0)$ and 
triplet $(S=1)$ as a function of magnetic field.  The spin singlet and 
triplet states correspond to even and odd angular momentum 
quantum numbers, $m$. 
\label{fig1}}
\end{figure}

Fig.~\ref{fig1}  plots 
$E_m-E_{\textrm{ground}}$ versus B for the 
four lowest energy states, $m=1,2,3,$ and $4$.  Cusps appear 
at $E_m-E_{\textrm{ground}}=0$ where the ground 
state changes at $B_m$ signaling a change in the number of 
vortices per electron.  (Note that the relation for $B_m$ is valid for 
$\omega_0 \ll \omega_c$.) 
The ground state clearly shows a number of 
spin transitions with increasing magnetic field \cite{Wagner}.  
Furthermore, the second highest excited state 
becomes degenerate with the third at level crossings which 
occur at magnetic fields between ground state transitions. 
This suggests that quantum information stored in the two lowest energy 
spin states in neighboring quantum dots becomes susceptible to leakage 
when the dots are brought very close together.  Below, we address this issue 
quantitatively using both exact diagonalization and a new set of 
variational states.  
   
We now diagonalize the full Hamiltonian, Eq.~(\ref{FullH}), in regimes
not amenable to perturbation theory, for $R\sim a$ and $\omega_0 
\lesssim \omega_c$.  
We construct the matrix
representing $H$ in the Fock-Darwin \cite{FockDarwin} basis 
centered between the two dots.  Previous studies have 
involved diagonalization of 
similar Hamiltonians using several dot centered basis states.  This 
technique requires  lengthy, numerical routines to 
generate an orthogonal set of basis states \cite{Hu,Loss2}. 
The limited number of basis states allows for high accuracy 
only in a regime where
the Coulomb interaction may be treated perturbatively.  However, 
to access the strongly correlated regime, we find 
it necessary to include up to $\sim 10^5$ origin centered, Fock-Darwin 
basis states with $z$-component of angular momentum 
less than twelve.  We use 
a modified Lanczos routine to obtain the ground and excited states. 
This technique yields the $entire$ spectrum.  However, here we focus 
on the four lowest energy states. 
The energies converge to within $1 \mu$eV upon inclusion of more 
basis states and may therefore be considered exact.  
Details, including the matrix elements, will be 
published elsewhere.  

\begin{figure}
\includegraphics[width=2.6in]{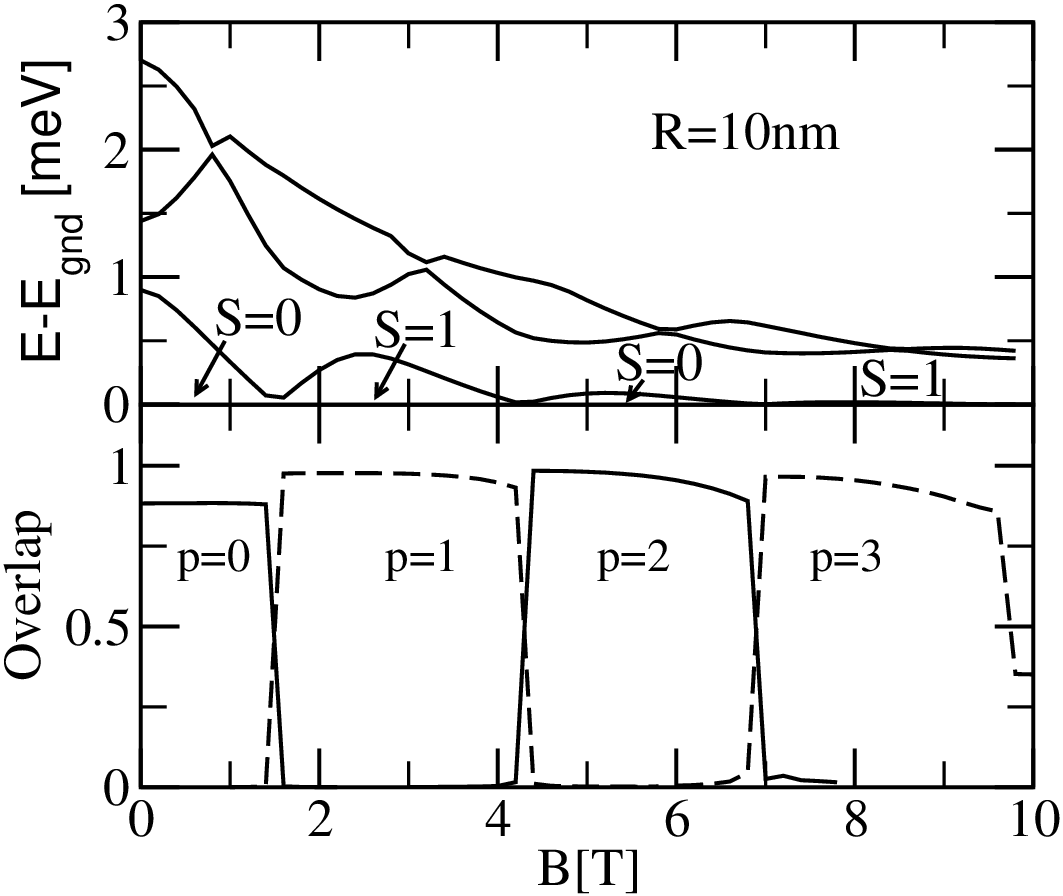}
\caption{
The top panel plots the energy of the four lowest states 
obtained by exact diagonalization of Eq.~(\ref{FullH}) as a function 
of magnetic field with the ground state energy set to zero.  
The energies converge to within $1 \mu$eV.  The 
separation between parabolic dots is $R=10$ nm.  Transitions between 
spin singlet and triplet states remain.  The bottom panel shows the overlap of 
the exact ground state and the trial states given by
Eq.~(\ref{states}).  The number of vortices attached to each electron 
increases with magnetic field from $p=0$ to $p=3$.  As in
Fig.~\ref{fig1}, singlet (triplet) states correspond to even (odd) 
values of $p$.  
\label{fig2}}
\end{figure}

The top panel in Fig.~\ref{fig2} shows the  
four lowest energies obtained from exact diagonalization of 
Eq.~(\ref{FullH}) versus magnetic field.  The energy zero is taken to
be the ground state.  We have chosen an inter-dot separation of
$R=10$ nm.  
The energy of the 
first excited state gives the effective exchange splitting  
which changes sign through successive spin transitions 
at each cusp.  The results are qualitatively similar to the results 
shown in Fig.~\ref{fig1} but are entirely unexpected.  
Vortex attachment non-perturbatively lowers the Coulomb energy 
of uniform states but does not necessarily apply to 
highly disordered systems.  Yet, the intriguing oscillations in the effective 
exchange interaction seen in Fig.~\ref{fig2} suggest just this 
and therefore require further study.

In comparing Figs.~\ref{fig1} and ~\ref{fig2} we find further 
differences. 
At low fields, the top panel of Fig.~\ref{fig2} correctly shows 
a spin singlet ground state 
at $B=0$ rather than a triplet state as shown in the 
LLL limit of Fig.~\ref{fig1}.  Most importantly the 
degeneracies in excited states at $B=0, 2.4, 5.2$ and $8$ T 
begin to lift.  As opposed to the level crossing in 
the single dot, $R=0$ case discussed earlier, 
the breaking of rotational symmetry forces an
anticrossing among the first and second excited states.  
Where, at small to intermediated inter-dot separations, $R\lesssim a$,  
the higher excited states 
are perturbed, single dot states with a nearly uniform charge density.
A large anticrossing among the two lowest excited 
states protects the quantum 
information stored in the entangled state of two strongly coupled 
quantum dots.  Experimental uncertainties in $R$ and $\omega_0$ may  
eventually lead to the strongly coupled regime.  Careful study 
of the states making up the anticrossing is therefore crucial.

We now construct variational states which reproduce the exact results 
discussed above.  The composite fermion theory \cite{Jain} of the 
fractional quantum Hall effect offers an accurate variational ansatz
describing two dimensional electron systems at high magnetic field.
A composite fermion is the bound state of an electron and an even
number of quantum mechanical vortices of the many-body wave 
function.  The corresponding 
orbital wave function is \cite{Jain}: $\psi=J\phi$, where 
$\phi$ is a weakly interacting fermion state, $J$ a Jastrow factor,
and $\psi$ the highly correlated state of electrons.  
In isotropic, spinless systems $J$ attaches 
an even number of vortices to the fermions in the antisymmetric 
state $\phi$ yielding an antisymmetric electron 
wave function.  In anisotropic systems with 
additional quantum numbers one may bind 
an even $or$ odd number of vortices to 
each particle while preserving the antisymmetry of the 
overall wave function \cite{Fertig,Scarola2}. 
Applying the composite fermion ansatz to the Hamiltonian studied 
here we take $\phi$ to 
be the non-interacting ground state. We also 
take $J=\left(z_1-z_2\right)^p$, where $p=0,1,2,..$, giving:
\begin{eqnarray}  
\overline{\psi}_p=(z_1-z_2)^p\phi.
\label{states}
\end{eqnarray}
This is our initial, high field solution of  
$H$.  $\overline{\psi}_p$ clearly reduces 
to $\vert m\rangle$ at $R=0$.  For $R\gg a$ the fermions in the 
state $\phi$ become localized on each dot leaving  
$J$ constant $\sim R^p$.  In which case 
$\overline{\psi}_p$ reduces to the limit of 
two independent electrons.  

The bottom panel of Fig.~\ref{fig2} plots the overlap of the 
exact ground state and the variational state
$\overline{\psi}_p$ at $R=10$ nm.  
Triplet (singlet) spin states correspond to odd (even) values of 
$p$, as in the $R=0$ case.  The overlaps drop to zero when the 
particle exchange symmetry of the orbital wave function changes. 
We have checked by direct calculation of the density that, 
by $B\sim 9$ T, the modified magnetic length has become small 
enough to localize the electrons on each dot. 
The surprisingly high overlaps prove that vortex
attachment is a valid ansatz even in the highly localized 
regime.  At large dot separations, $R\geq 40$ nm, the Coulomb
interaction lowers to a point where the splitting 
between spin states is near zero at large B.  However, we 
have checked that even here the 
overlaps remain large.  
Another important feature of $\overline{\psi}_p$
is that the $p=0$ state does not take into 
account the Coulomb interaction.  The 
overlaps near $B=0$ are correspondingly lower.  

To improve the variational states in the low field 
regime, $\omega_0 \geq \omega_c$, we note that the 
Coulomb energy cost maybe lowered by mixing with higher 
energy states of the quantum dot which serves to 
increase the average inter-electron separation.  In the 
$R=0$ case, rotational symmetry requires the addition  
of states with the same angular momentum.
This leads to the following trial states:
\begin{eqnarray} 
\psi_p=(z_1-z_2)^p\left(1
+\beta b^\dagger a^\dagger \right)\phi,
\label{general}
\end{eqnarray}
where the variational parameter $\beta$ 
controls the amount of mixing with higher 
energy levels of the quantum dot.
The total raising operators 
$b^\dagger=b^\dagger_1+b^\dagger_2$ and 
$a^\dagger=a^\dagger_1+a^\dagger_2$ act on the 
Fock-Darwin basis states.  The single particle 
raising operators are given by: 
$b_j^\dagger=(z_j^*/2-2\partial_{z_j})/\sqrt{2}$
and
$a^\dagger_j=i(z_j/2-2\partial_{z_j^*})/\sqrt{2}$.
The above variational states include mixing with 
higher energy states of the same angular momentum because the 
operator $b^\dagger a^\dagger$ does not change the angular momentum 
of a Fock-Darwin state.  

\begin{figure}
\includegraphics[width=2.3in]{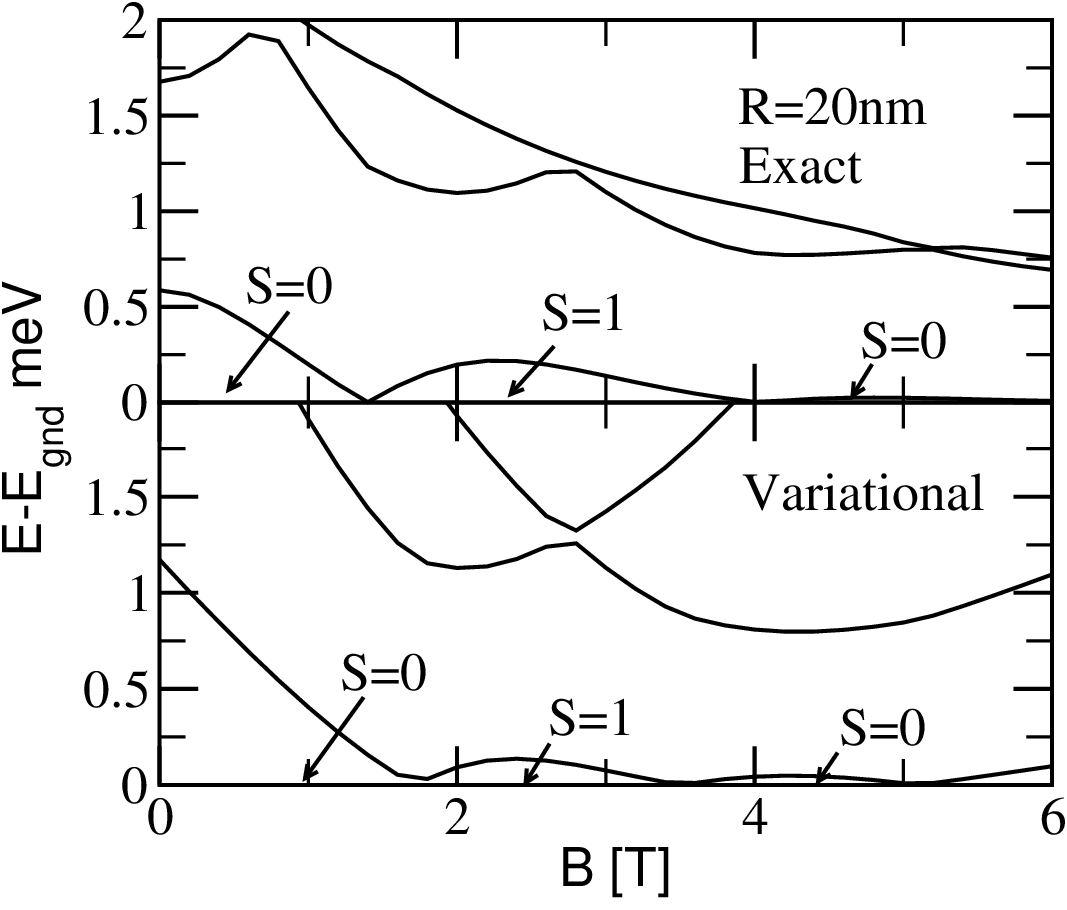}
\caption{  
The top panel shows the same as Fig.~\ref{fig2} but for a dot
separation of $R=20$ nm.  The bottom panel shows the energy of 
the trial states in Eq.~(\ref{general}) 
as a function of magnetic field.  The ground state energy 
is set to zero.  The energies are obtained by 
orthogonalizing the four variational states with 
$p=0, 1, 2,$ and $3$ and diagonalizing Eq.~(\ref{FullH}).  The variational 
parameter $\beta$ is chosen to minimize the total energy.  
\label{fig3}}
\end{figure}

The top panel in Fig.~\ref{fig3} plots the exact energy spectrum, as in
Fig.~\ref{fig2}, but for $R=20$ nm.  Here we see that, at large
magnetic fields, the large separation between electrons localized on 
each dot suppresses the exchange splitting.  However, several 
spin transitions still remain.  The bottom panel in Fig.~\ref{fig3}
shows the energy of the four variational states 
$\psi_p$, with $p=0, 1, 2,$ and $3$.  We take the ground state 
to be the zero in energy.  We
obtain the energy by orthogonalizing the four 
variational states and diagonalizing $H$ in this four state 
basis.  We minimize the energy with respect to $\beta$ at each $B$. 
The parameter 
$\beta$, of the ground state varies from 0.02 at $B=0$ 
to 0.0006 at $B=5$ T showing that 
large magnetic fields all but suppress Landau level mixing.  
The exchange splitting obtained 
with the variational states compares well with the exact value.  
Furthermore, in the range $B=1$ to $4$ T, the second excited state 
captures the essential features of the corresponding exact results.
Rotational symmetry breaking forces the higher excited states 
to open an anticrossing observed near 
$B=0,2.4,$ and $4.3$ T.
The states at the anticrossings in Fig.~\ref{fig3} are 
similar to the states making up the level crossings 
in Fig.~\ref{fig1}.  For example the electrons in 
the first excited state at $2.4$ T in Fig.~\ref{fig3} 
form a two and zero vortex mixed state in a   
56\% to 44\% ratio, as opposed to the ground state 
which holds one vortex per electron, to within 98\%.
To evaluate the anticrossing 
explicitly we note that for 
$R\ll a$ the asymmetry in confinement acts as a perturbation.  We 
may rewrite the confinement potential up to an overall constant:
\begin{eqnarray}
V(\omega_0,R;\textbf r)=\frac{m^*\omega_0^2}{2}\left(\vert\textbf{r}\vert^2-\vert x\vert R\right).
\end{eqnarray}  
The second term breaks rotational symmetry and forces an anticrossing 
among the lowest two excited states.  It is crucial that the 
two lowest excited states involve states of even vorticity.  
Symmetry allows these two states to mix yielding an 
anticrossing as one may find by diagonalizing the 
rotational symmetry breaking term in the even-vorticity 
subspace.  The matrix elements are:
$m^*\omega_0^2R/2\left\langle\psi_{p'}\vert \vert 
x_1 \vert + \vert 
x_2 \vert \vert \psi_{p}\right\rangle$, where, 
near $B=2.4$ T for example, 
$p$ and $p'$ may be 0 or 2.  These matrix elements give an 
anticrossing $\sim m^*\omega_0^2Ra$.  This is 
in contrast to ground state transitions between 
states with even and odd vorticity.  Here the states $\psi_p$ and 
$\psi_{p+1}$ cannot mix, allowing the exchange splitting to 
change sign.

We stress that the top panel in Fig.~\ref{fig3} is obtained 
by diagonalization of the full 
Hamiltonian with $\sim 10^5$ basis states while the lower panel is 
obtained by the same method but with four, physically relevant  
basis states.  The 
agreement breaks down at larger fields, $B\sim 5.6$ T, where the
excited states should include the $p=4$ variational state.  Inclusion 
of variational states with large $p$ is necessary at larger fields.  
The excellent agreement obtained thus far demonstrates that the
plethora of spin transitions in strongly coupled quantum dots originates 
from a swapping of the particle exchange symmetry associated 
with vortex attachment.

To conclude, we have discussed the surprisingly rich 
and unexpected vortex structure found in the exchange coupling 
of a two-electron artificial molecule in coupled semiconductor 
quantum dots which should be important both for constructing 
quantum gates and for studying strongly correlated quantum dot 
electronic states.

We thank J.K. Jain and K. Park for helpful discussions.



\end{document}